\documentclass[onecolumn,showpacs]{revtex4}

\usepackage{graphicx}
\usepackage{dcolumn}
\usepackage{bm}

\input epsf

\begin{document}

\title{\Large Does Cosmic No-Hair Conjecture in Brane Scenarios follow from
General Relativity?}

\author{\bf Subenoy Chakraborty}
\email{subenoyc@yahoo.co.in}
\author{\bf Ujjal Debnath}
\email{ujjaldebnath@yahoo.com}
\affiliation{Department of
Mathematics, Jadavpur University, Calcutta-32, India.}

\date{\today}

\begin{abstract}
In this paper we examine the Cosmic No-Hair Conjecture (CNHC) in
brane world scenarios. For the validity of this conjecture, in
addition to the strong and weak energy conditions for the matter
field, a similar type of assumption is to be made on the
quadratic correction term and there is a restriction on the
non-local term. It is shown by examples with realistic fluid
models that strong and weak energy conditions are sufficient for
CNHC in brane world.
\end{abstract}


\maketitle

\section{\normalsize\bf{Introduction}}
The idea of brane world scenarios may resolve the challenging
problem in theoretical physics namely the unification of all
forces and particles in nature. It is suggested that we live in a
four dimensional brane embedded in a higher dimensional
space-time. As a result, the fundamental higher dimensional
Planck mass could be of same order as the electro weak scale and
thereby one of the hierarchy problems in the current standard
model of high-energy physics are resolved [1-4].\\

According to Randall and Sundrum [3, 4] it is possible to have a
single massless bound state confined to a domain wall or 3-brane
in five dimensional non-factorizable geometries. They have shown
that this bound state corresponds to the zero mode of the
Kaluza-Klein dimensional reduction and is related to the four
dimensional gravitation [2]. Hence all matter and gauge fields
(except gravity) are confined to a 3-brane embedded in a five
dimensional space-time (bulk) while gravity can propagate in the
bulk. As a consequence, the gravity on the brane can be described
by the Einstein's equations modified by two additional terms,
namely (i) quadratic in matter variables and (ii) the electric
part of the five dimensional Weyl tensor [5].\\

In general terminology, the CNHC [6, 7] states that ``all
expanding universe models with a positive cosmological constant
asymptotically approach the de-Sitter solution". To address the
question whether the universe evolves to a homogeneous and
isotropic state during an inflationary epoch, Gibbons and Hawking
[6] and then Hawking and Moss [7] developed this conjecture.
Subsequently, Wald [8] gave a formal proof of it for homogeneous
cosmological models (Bianchi models) with a positive cosmological
constant. He assumed that the matter field should satisfy strong
and weak energy conditions.\\

In this paper we wish to extend Wald's [8] result in the brane
world scenario and examine whether the new conditions can be
minimize using those in general relativity.

\section{\normalsize\bf{Cosmic No Hair Conjecture in Brane World}}
According to Roy Maartens [1], the Einstein equations on the
brane can be written as

\begin{equation}
G_{\mu\nu}=-\Lambda
g_{\mu\nu}+\kappa_{4}^{2}T_{\mu\nu}+\kappa_{5}^{4}S_{\mu\nu}-E_{\mu\nu}
\end{equation}

where $S_{\mu\nu}$ and $T_{\mu\nu}$ are the two correction terms
(local and non-local) in the energy momentum tensor. The local
correction term $S_{\mu\nu}$ has the expression

\begin{equation}
4S_{\mu\nu}=\frac{1}{3}T
T_{\mu\nu}-T_{\mu\rho}T_{\nu}^{\rho}-\frac{1}{2}g_{\mu\nu}\left(\frac{1}{3}T^{2}
-T_{\rho\sigma}T^{\rho\sigma}\right)
\end{equation}

while $E_{\mu\nu}$ is the electric part of the 5D Weyl tensor in
the bulk. Now the scalar constraint (initial value constraint)
equation and the Roychoudhuri equation on the brane has the form

\begin{equation}
G_{\mu\nu}n^{\mu}n^{\nu}=\Lambda+\kappa_{4}^{2}T_{\mu\nu}n^{\mu}n^{\nu}
+\kappa_{5}^{4}S_{\mu\nu}n^{\mu}n^{\nu}-E_{\mu\nu}n^{\mu}n^{\nu}
\end{equation}

and

\begin{equation}
R_{\mu\nu}n^{\mu}n^{\nu}=-\Lambda+\kappa_{4}^{2}\left(T_{\mu\nu}-\frac{1}{2}
g_{\mu\nu}T\right)n^{\mu}n^{\nu}+\kappa_{5}^{4}\left(S_{\mu\nu}-\frac{1}{2}
g_{\mu\nu}S\right)n^{\mu}n^{\nu}-E_{\mu\nu}n^{\mu}n^{\nu}
\end{equation}

where $n^{\mu}$ is the unit normal to the spatial homogeneous
hypersurfaces. In terms of the homogeneous hypersurface elements
namely, the projected metric
$h_{\mu\nu}(=g_{\mu\nu}+n_{\mu}n_{\nu})$ and the extrinsic
curvature $K_{\mu\nu}(=\nabla_{\nu}n_{\mu})$ and using the
Gauss-Codazzi equations the above two equations namely equations
(3) and (4) become

\begin{equation}
K^{2}=3\Lambda+\frac{3}{2}\sigma_{\mu\nu}\sigma^{\mu\nu}-\frac{3}{2}~^{(3)}R
+3\kappa_{4}^{2}T_{\mu\nu}n^{\mu}n^{\nu}+3\kappa_{5}^{4}S_{\mu\nu}n^{\mu}n^{\nu}
-3E_{\mu\nu}n^{\mu}n^{\nu}
\end{equation}

and

\begin{equation}
\dot{K}=\Lambda-\frac{1}{3}K^{2}-\sigma_{\mu\nu}\sigma^{\mu\nu}-
\kappa_{4}^{2}\left(T_{\mu\nu}-\frac{1}{2}g_{\mu\nu}T\right)n^{\mu}n^{\nu}
-\kappa_{5}^{4}\left(S_{\mu\nu}-\frac{1}{2}g_{\mu\nu}S\right)n^{\mu}n^{\nu}
+E_{\mu\nu}n^{\mu}n^{\nu}
\end{equation}

where the dot denotes the Lie derivative with respect to proper
time, $K$ is the trace of the extrinsic curvature,
$\sigma_{\mu\nu}$ is the shear of the time like geodesic
congruence orthogonal to the homogeneous hypersurfaces and
~$^{(3)}R$ is the scalar
curvature of the homogeneous hypersurfaces.\\

Using the idea of Wald and proceeding along his approach (for
details see Wald et al [8] and Chakraborty et al [9]) one can find
that for CNHC we must have

\begin{equation}
(a)~~~~~~~~~~~~~~~~~~~~~ S_{\mu\nu}n^{\mu}n^{\nu} \ge 0 ~~~~
\text{and}
~~~~\left(S_{\mu\nu}-\frac{1}{2}g_{\mu\nu}S\right)n^{\mu}n^{\nu}
\ge 0 \hspace{1in}
\end{equation}

and

\begin{equation}
(b)~~~~~~~~~~~~~~~~~~~~~~~~~~~~~~~~~~~~~~~~~~~~~~E_{\mu\nu}n^{\mu}n^{\nu}
\le 0 \hspace{2in}
\end{equation}

in addition to the weak and strong energy conditions for the
matter field

\begin{equation}
T_{\mu\nu}n^{\mu}n^{\nu} \ge 0
~~~\text{and}~~~\left(T_{\mu\nu}-\frac{1}{2}g_{\mu\nu}T\right)n^{\mu}n^{\nu}
\ge 0
\end{equation}

Now if we use the expression (2) for $S_{\mu\nu}$ in (7) then we
get

\begin{equation}
\frac{1}{3}T
b-\frac{1}{2}T_{\rho\sigma}T^{\rho\sigma}-(T_{\mu\rho}n^{\mu})(T_{\nu}^{\rho}n^{\nu})
\ge 0
\end{equation}

and

\begin{equation}
\frac{1}{3}T a-(T_{\mu\rho}n^{\mu})(T_{\nu}^{\rho}n^{\nu}) \ge 0
\end{equation}

where $a=T_{\mu\nu}n^{\mu}n^{\nu}$ and
$b=\left(T_{\mu\nu}-\frac{1}{2}g_{\mu\nu}T\right)n^{\mu}n^{\nu}$
are positive due to (9).\\

Also using the symmetry properties of $E_{\mu\nu}$, it is
possible to decompose it with respect to any time like observer
$\vec{u}~(u^{\alpha}u_{\alpha}=-1)$ as [5]

$$
E_{\mu\nu}=-\left(\frac{\kappa_{5}}{\kappa_{4}}\right)^{4}\left[\left(u_{\mu}u_{\nu}+
\frac{1}{3}h_{\mu\nu}\right)U+2(u_{\mu}Q_{\nu})+P_{\mu\nu}\right]
$$

with the properties

$$
Q_{\mu}u^{\mu}=0,~~ P_{(\mu\nu)}=P_{\mu\nu},~~ P_{\mu}^{\mu}=0,~~
P_{\mu\nu}u^{\nu}=0
$$

If we consider the Bianchi models then due to the symmetry of the
spatial geometry we may choose

$$
Q_{\mu}=P_{\mu\nu}=0,
$$

and the scalar part namely $U$ is termed as dark energy density as
it has energy-momentum tensor that of a radiation perfect fluid.
So the restriction (8) implies that dark energy density should be
always positive i.e.,

\begin{equation}
U\ge 0
\end{equation}

As it is not possible to make any restriction on $T_{\mu\nu}$ to
satisfy inequations (10) and (11), so let us examine with some
realistic model for the matter field.

\section{\normalsize\bf{Examples}}
$$
\text{\bf(a)~~~Perfect fluid model:}
$$

In this case the energy-momentum tensor has the form

$$
T_{\mu\nu}=(\rho+p)n_{\mu}n_{\nu}+p
g_{\mu\nu},~~~n_{\mu}n^{\mu}=-1
$$

with $\rho$ and $p$ as the energy density and isotropic pressure
respectively.\\

The weak and strong energy conditions demand

\begin{equation}
a=\rho \ge 0 ~~~\text{and}~~~ 2b=\rho+3p \ge 0
\end{equation}

Hence the inequations (7) (i.e., inequations (10) and (11)) take
the form

\begin{equation}
\rho^{2} \ge 0 ~~~\text{and} ~~~ \rho(3p+2\rho) \ge 0
\end{equation}

which are always true. Thus for perfect fluid model CNHC is
automatically satisfied in brane scenarios if it is valid in
general relativity.
$$
\text{\bf (b)~~~General form of energy-momentum tensor:}
$$

The general form of the brane energy momentum tensor for any
matter fields (scalar field, perfect fluids, kinematic gases,
dissipative fluids, etc.) including a combination of different
fields can be covariently written as [1]

\begin{equation}
T_{\mu\nu}=\rho n_{\mu}n_{\nu}+p
h_{\mu\nu}+\Pi_{\mu\nu}+q_{\mu}n_{\nu}+q_{\nu}n_{\mu}
\end{equation}

Here the energy flux $q_{\mu}$ and the anisotropic stress
$\Pi_{\mu\nu}$ are projected, symmetric and traceless that is
$$
q_{\mu}n^{\mu}=0,~~\Pi_{\mu\nu}n^{\mu}=0,~~\Pi_{\mu\nu}=\Pi_{\nu\mu},
~~\Pi_{\mu\nu}g^{\mu\nu}=0
$$

Thus for this form of energy-momentum tensor, the restrictions on
$S_{\mu\nu}$ now result

\begin{equation}
\frac{1}{3}\rho^{2}-\frac{1}{2}\Pi_{\mu\nu}\Pi^{\mu\nu} \ge 0
\end{equation}

and

\begin{equation}
\frac{1}{3}\rho(3p+2\rho)-q_{\mu}q^{\mu} \ge 0
\end{equation}

As for realistic matter, the energy density should be larger than
the anisotropic stress and heat flux is very small in magnitude
so the inequalities (16) and (17) are automatically satisfied.
Hence the CNHC is satisfied for the above form of general
energy-momentum tensor.\\

For future work, it will be interesting to find any general
restrictions on $T_{\mu\nu}$ so that CNHC is automatically
satisfied in brane world scenario.\\

{\bf Acknowledgement:}\\

The authors are thankful to Prof. U. C. De of Kalyani University
for valuable discussion. The authors are also thankful to D.
Marolf for his valuable suggestions. Finally the authors are
grateful to referees' for their comments which improve the paper.
One of the authors (U.D) is thankful to CSIR (Govt. of India)
for awarding a Junior Research Fellowship.\\

{\bf References:}\\
\\
$[1]$  Maartens Roy, Reference Frames and Gravitomagnetism, eds J
F Pascual-Sanchez, L Floria, A San Miguel, F Vicente (World
Scientific, 2001) pp. 93-119;
(also {\it gr-qc}/0101059 (2001)).\\
$[2]$  Chakraborty S and Chakraborti S, {\it Class. Quantum Grav.}
{\bf 19} 3775 (2002).\\
$[3]$  Randall L and Sundrum R, {\it Phys. Rev. Lett.} {\bf 83}
3770 (1999).\\
$[4]$  Randall L and Sundrum R, {\it Phys. Rev. Lett.} {\bf 83}
4690 (1999).\\
$[5]$  Maartens R, {\it Phys. Rev. D} {\bf 62} 084023 (2000);
Campos A and Sopuerta C F, {\it Phys. Rev. D} {\bf 63} 404012 (2001);
{\it Phys. Rev. D} {\bf 64} 104011 (2001).\\
$[6]$  Gibbons G W and Hawking S W, {\it Phys. Rev. D} {\bf 15} 2738 (1977).\\
$[7]$  Hawking S W and Moss I G, {\it Phys. Lett.} {\bf 110B} 35 (1982).\\
$[8]$  Wald R M, {\it Phys. Rev. D} {\bf 28} 2118 (1983).\\
$[9]$  Chakraborty S and Paul B C, {\it Phys. Rev. D} {\bf 64}
127502 (2001);

\end{document}